\documentclass[10pt,journal,compsoc]{IEEEtran}
\newcommand{\etal}{et~al.~} 
\newcommand{\ie}{i.e.,~}
\newcommand{\eg}{e.g.,~}

\usepackage{microtype}                 
\PassOptionsToPackage{warn}{textcomp}  
\usepackage{textcomp}                  
\usepackage{mathptmx}                  
\usepackage{times}                     
\usepackage{tabu}                      
\usepackage{booktabs}                  
\usepackage{enumitem}
\usepackage{url}
\makeatletter
\g@addto@macro{\UrlBreaks}{\UrlOrds}
\makeatother
\usepackage{graphicx}
\usepackage{wrapfig}
\usepackage[T1]{fontenc}
\usepackage{amsmath}  
\usepackage{xcolor}  
\newcommand{\x}[1]{\textcolor{black}{#1}}
\usepackage{multirow}
\usepackage{tabularx}
\usepackage{soul}

\ifCLASSOPTIONcompsoc
  \usepackage[nocompress]{cite}
\else
  \usepackage{cite}
\fi
\ifCLASSINFOpdf
\else
\fi
\begin{document}
%
\title{\x{The Chart Excites Me! Exploring How Data Visualization Design Influences Affective Arousal}}
%
%
%
%

\author{
Xingyu Lan, Yanqiu Wu, Qing Chen, and Nan Cao
\IEEEcompsocitemizethanks{\IEEEcompsocthanksitem Xingyu Lan, Yanqiu Wu, Qing chen, and Nan Cao are with
Intelligent Big Data Visualization Lab at Tongji University.\protect\\
E-mail: \{xingyulan96, wuyanqiu.idvx, jane.qing.chen, nan.cao\}@gmail.com. Nan Cao is the corresponding author.}
}

\IEEEtitleabstractindextext{%
\begin{abstract}
\x{As data visualizations have been increasingly applied in mass communication, designers often seek to grasp viewers immediately and motivate them to read more. Such goals, as suggested by previous research, are closely associated with the activation of emotion, namely affective arousal.}
Given this motivation, this work takes initial steps toward understanding the arousal-related factors in data visualization design. We collected a corpus of 265 data visualizations and conducted a crowdsourcing study with 184 participants during which the participants were asked to rate the affective arousal elicited by data visualization design (all texts were blurred to exclude the influence of semantics) and provide their reasons. \x{Based on the collected data, first, we identified a set of arousal-related design features by analyzing user comments qualitatively. Then, we mapped these features to computable variables and constructed regression models to infer which features are significant contributors to affective arousal quantitatively. Through this exploratory study, we finally identified four design features (\eg colorfulness, the number of different visual channels) cross-validated as important features correlated with affective arousal.
} 
\end{abstract}

}

\maketitle

\IEEEdisplaynontitleabstractindextext

%
\IEEEpeerreviewmaketitle

\ifCLASSOPTIONcompsoc
\IEEEraisesectionheading{\section{Introduction}\label{sec:introduction}}
\else
\section{Introduction}
\label{sec:introduction}
\fi

\IEEEPARstart{D}{\x{ata}} \x{visualization is now widely used in mass communication in forms such as infographics, data-driven articles, and data videos~\cite{lan2021smile,segel2010narrative,shi2021communicating}.
Due to the traits of mass communication (\eg information overload, impatient viewers), visualization designers often need to activate viewers emotionally to grasp them immediately and motivate them to read more.
In research, this design goal is closely related to a concept called affective arousal, which measures the activation of emotion on a continuous scale~\cite{russell1980circumplex}.
Research from fields such as marketing and communications has provided abundant evidence that affective arousal can moderate people's willingness to spend time and money on visual media~\cite{sherman1997store,kaltcheva2006should}or lead to behaviors such as clicking and sharing~\cite{berger2011arousal}, thereby making it an important perspective of investigating visual communication~\cite{storbeck2008affective}.
In the visualization community, some practitioners have also stressed the importance of affective arousal. 
For example, a data visualization that looks exciting is usually more desirable in advertising as it can attract more clicks, likes, and sharing~\cite{lin2019minimal}. 
For nonprofits, designers may want a data visualization to elicit high-arousal affects in order to raise funds and call for social collaboration~\cite{chibana2015nonprofit}. }
In addition, previous work has shown that design elements such as color and imagery can effectively influence arousal levels~\cite{valdez1994effects,storbeck2008affective}, which from another perspective confirms the possibility of conducting research on the relationship between data visualization design and affective arousal.


Although many researchers have acknowledged that affects are an indispensable part of users' experience with data~\cite{wang2019emotional,bartram2017affective,lan2021smile}, previous literature has mostly viewed affects as discrete categories rather than examining them in a continuous space.
For example, Kennedy and Hill~\cite{kennedy2018feeling} investigated people's affective responses to data visualizations by conducting a diary study. The affects reported by the participants were mostly categorical, such as surprise and disgust. Boy~\etal~\cite{boy2017showing} focused on examining how data visualization design influences one particular type of affect, empathy. Lan~\etal~\cite{lan2021kineticharts} proposed an affective animation design scheme for common charts to convey positive affects such as joy and tenderness in data stories. 
\x{In their recent work about serious data storytelling~\cite{lan2022negative}, the researchers focused on examining how design methods facilitate the communication of negative emotions (\eg anxiety, helplessness).
The work by Bartram~\etal~\cite{bartram2017affective} is more relevant to affective arousal. They proposed a set of color palettes to help encode eight types of affect that represent typical combinations of valence and arousal (\eg positive, negative, calm, exciting).}
However, this study still simplified affects as categories and did not show how affects change in the dimensional space continuously.
To sum up, although the above work has contributed to our understanding of how data visualization design influences affects, as yet no systematic work has been carried out to address this problem through the lens of affective arousal.

To fill this gap, this study first explores what data visualization design factors may influence the perception of affective arousal. To begin with, we prepared a corpus containing 265 data visualization images based on the MASSVIS dataset~\cite{borkin2015beyond}, \x{which provides single visualizations from balanced sources such as infographics, news media, and governments. To prevent the bias caused by semantics (because the content of the images can strongly influence affects), we blurred all the texts on these images.} Based on this corpus, we conducted a crowdsourcing study involving 184 participants and asked the participants to assess the affective arousal elicited by the designs in the corpus. We then analyzed the experimental results and derived a set of design features that the participants thought related to affective arousal by coding their comments manually. These design features were grouped into three main categories: color, graphics, and information.
\x{Next, we mapped these features to computable variables and constructed regression models to identify the most significant features and assess their importance.
By comparing the results of the models, we finally identified four design features that were cross-validated as important features correlated with affective arousal, including colorfulness, the number of different visual channels, the X and Y coordinates, and the grid layout.}
Finally, based on all the quantitative and qualitative analyses above, we discuss the implications that arise from our work, reflect on the limitations of this work, and propose future research opportunities.



\section{Related Work}

\subsection{\x{Data Visualization for Communication}}

\x{In recent years, a notable phenomenon in the visualization community is that data visualization has been massively applied in mass communication 
(\ie the process of creating, sending, imparting, and exchanging messages to a large number of audiences) 
through media such as infographics, data-driven articles, and data videos~\cite{shi2020calliope,lan2021smile,shi2021communicating}. 
However, on the other hand, the features of contemporary mass communication have posed new challenges for visualization design. 
As found by research in communications~\etal~\cite{lorenz2019accelerating}, the abundance of information produced by mass media today has narrowed people's attention span and made them more and more impatient.
To grasp and engage more viewers, designers have to consider user experience and modify their design according to how viewers might respond.
}

\x{In response to such challenges, a growing number of studies in the visualization community have adopted a user-centered perspective and examined visualization design through the lens of user experience.
For example, Saket~\etal~\cite{saket2016beyond} argued that more efforts should be put into understanding people's subjective responses to visualization design.
Cawthon~\etal~\cite{cawthon2007effect} and Harrison~\etal~\cite{harrison2015infographic} investigated how data visualization influences aesthetic feelings. Amini~\etal~\cite{amini2018hooked} and Lan~\etal~\cite{lan2021understanding} used indicators such as likability and enjoyment to evaluate data stories. 
Borkin~\etal~\cite{borkin2013makes} evaluated a series of visualization designs by assessing how memorable they were.
However, so far, existing studies have only addressed limited aspects of user experience. Many important concepts are left underexplored.}

\x{Therefore, in this work, we choose to focus on affect arousal, a core construct of user experience in mass communication, and examine how it correlates with data visualization design. 
We hope our work can contribute a new perspective to understanding data communication and user-centered visualization design.}

\subsection{Affective Arousal}

Affective arousal is about the activation of emotion~\cite{russell1980circumplex} and is viewed as a key construct of human affects. 
For example, the well-known circumplex model~\cite{russell1980circumplex} views affects as combinations of valence and arousal. While valence describes whether an affect is positive or negative, arousal reflects the intensity of the affect, ranging from calm to excitement. 
\x{Along with the conceptualization of affective arousal, many researchers began to conduct studies to investigate how to elicit affective arousal specifically. For example, Sundar and Kalyanaraman~\cite{sundar2004arousal} examined the relationship between animation and affective arousal using physiological indicators. Lang~\etal~\cite{lang1999effects} and Gorn~\etal~\cite{gorn1997effects} used self-report methods to gather ratings from users and assess how color and pacing influenced perceived arousal.}

\x{Meanwhile, as affective arousal is closely related to people's reactions, behaviors, and information processing~\cite{storbeck2008affective}, it
has been massively studied in research about visual communication, such as web advertising and TV broadcasting.
For example, Singh and Hitchon~\cite{singh1989intensifying} reviewed dozens of studies conducted in the last century that examined arousal-centered topics, such as how to arouse viewers most advantageously in a TV show and how to increase consumers' desire for a product by eliciting arousal.
Given the fundamental role of affective arousal, this thread of research continues to prosper until today while dealing with trending topics such as how to design arousing games, user interfaces, or immersive environments~\cite{sundar2004arousal,mavridou2018towards,cusveller2014evoking}. For example, Aoki~\etal~\cite{aoki2022emoballoon} examined how to design chat balloons that represent various levels of arousal in text chats. Cusveller~\etal~\cite{cusveller2014evoking} explored how to evoke affective arousal in computer games.
}
\x{This work follows this research thread and studies affective arousal specifically in the context of visualization design. We see this work as an initial step towards exploring what design factors contribute to affective arousal in data visualization. }

\subsection{Affective Design in Data Visualization}


In recent years, researchers~\cite{bartram2017affective,lan2021smile} have introduced the idea of \textit{affective design} to the community of data visualization and aimed to understand how design elements influence viewers' affective feelings towards data visualization.
For example, Bartram~\etal~\cite{bartram2017affective} identified a set of affective color palettes that can elicit eight categories of affects, such as calmness and positivity. 
Boy~\etal~\cite{boy2017showing} investigated one specific category of affects, empathy, and found that using anthropomorphic pictograms to represent human rights data does not help trigger empathy. 
Lan~\etal~\cite{lan2021smile} summarized a set of affect-related design factors in infographics, such as the usability of the design and the expressiveness of the design.
They also explored the dynamic design elements in data visualization design~\cite{lan2021kineticharts} and proposed an animation design scheme, Kineticharts, for creating charts that express five positive-valenced affects, such as joy, surprise, and amusement.
\x{After this work, they continued to explore negative-valenced affects in serious data storytelling~\cite{lan2022negative}.}

However, although the aforementioned work has laid a foundation for investigating how data visualization design influences affects, \x{it has two main limitations. First, previous work usually treats affects as discrete categories, and this paradigm has largely overlooked the continuous aspects of affects. Second, compared to other concepts such as valence, arousal has been less studied, thus limiting our understanding of how affects are activated by visualization design.}
To bridge this gap, this work investigates affective arousal specifically. As a continuous measurement, affective arousal helps capture the intensity of people's affective experience with visualization, thus contributing new knowledge around affective visualization design.

\section{Dataset and Pilot Study}

To identify design factors that contribute to affective arousal in data visualization design, we prepared a corpus of data visualization designs and conducted a crowdsourcing study with 184 participants. The participants were asked to view the data visualizations randomly chosen from the corpus and rate the affective arousal triggered by these images.
Afterward, we analyzed the data collected from the study to summarize a set of potential arousal-related design features. The study materials can be accessed at \url{https://bit.ly/3Ntv5ck}.

\subsection{Dataset}

We aimed to use a dataset that contains diverse data visualizations and reflects the variety of data visualization design in the real world. A dataset that meets such needs is the Massachusetts (Massive) Visualization Dataset (MASSVIS)~\cite{borkin2015beyond}. MASSVIS has collected thousands of data visualizations \x{by balancing four types of sources}: infographic galleries, news media, governments, and scientific journals. 
Due to its \x{representativeness in visualization design} and public availability, MASSVIS has been used frequently in previous studies~\cite{brehmer2016timelines,kim2021towards,borkin2013makes,borkin2015beyond}.
MASSVIS is constituted of several sub-corpora with different data sizes and labels. For example, \textit{single2k} is a corpus that contains more the 2000 pieces of single data visualizations labeled with basic meta-information such as sources and titles.
\textit{Targets410}, another sub-corpus, contains 410 representative data visualizations selected by experts from \textit{single2k}. It contains a more suitable amount of images to be used in a user study while maintaining the design diversity of \textit{single2k}.
Besides, \textit{targets410} has been labelled with more design-related tags, such as the visualization type and data-ink ratio. Therefore, \textit{targets410} suits our study better.

Besides, according to previous literature~\cite{lan2021smile,kennedy2018feeling}, the semantics carried by a data visualization will remarkably influence people's affects. Many people would react affectively to the content or the topic of a data visualization. Besides, embellishments such as an illustration of violence and a photograph of a child can also trigger strong affects. Such affects brought by semantics will be confounded with the affects brought by design and thus hinder our understanding of the exact impact of visualization design. Therefore, to get rid of the influence of semantics, we blurred all the texts and embellishments on the images so that the participants in our studies could focus on perceiving the affects triggered by design only.

\subsection{Pilot study}

We used the blurred version of \textit{targets410} as our stimuli and conducted a pilot study to investigate whether it was appropriate to use this corpus in our study. Based on the findings from the pilot study, we refined the corpus to construct the final stimuli for our crowdsourcing study.

\subsubsection{Methodology}

\begin{figure}[t]
 \centering
 \includegraphics[width=\columnwidth]{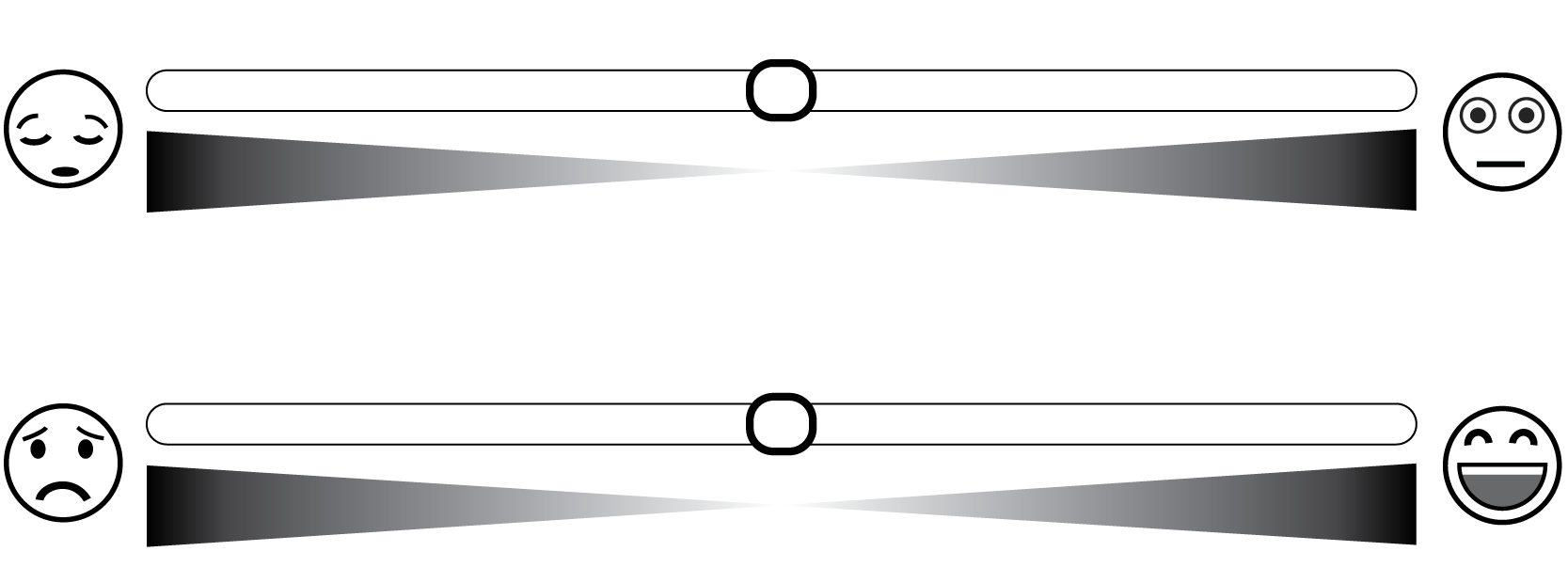}
 \caption{Affective Slider~\cite{betella2016affective} used in our study to rate the affective arousal (up) and valence (down) triggered by visualization design.}
 \label{fig:slider}
 \vspace{-1em}
\end{figure}

We recruited 10 participants (7 females, aging from 21 to 26) for the pilot study \x{by posting an open call on social media platforms}.
Once entering the online survey, the participants were presented with an introduction to our research, such as the definition of affective arousal and how to use the instrument for evaluating affects, namely Affective Slider (AS, see Fig.~\ref{fig:slider})~\cite{betella2016affective}. AS is a modern version of Self-Assessment Manikin (SAM)\cite{bradley1994measuring}, which is the traditional instrument for measuring valence and arousal based on the circumplex model. SAM adopts a 9-point scale (\eg for affective arousal, 1 denotes not at all aroused, and 9 denotes strongly aroused). 
AS inherits the core spirit of SAM but stretches the scale to 100 by providing a movable slider whose range is 0.00 to 1.00. Such a scale can produce more continuous and precise data, which is more suitable for use in statistical modeling.
Note that although this work focuses on arousal, we asked the participants to rate both arousal and valence during the study in case there was any interaction between these two dimensions.
After reading the introduction, the participants viewed 20 data visualizations randomly chosen from our corpus, one at a time. 
We also shuffled the order of the 20 stimuli to ensure the participants viewed a random sequence of images.
After viewing each of the images, the participants rated their arousal and valence using AS.
They were also asked to provide reasons for their ratings and rate the likability of the design (``Do you like this data visualization?'') using a 5-point Likert scale (1 denotes strongly dislike, 5 denotes strongly like). \x{The question about likability was included because prior literature has suggested that high arousal may not always be favorable~\cite{berlyne1960conflict}. Therefore, we followed their practice~\cite{gorn1997effects,lichtle2007effect} and evaluated whether the arousal elicited by stimuli was liked by the participants to facilitate the interpretation of arousal.}

After viewing and rating all 20 images, the participants filled out a form to answer four demographic questions (\ie gender, age, education level, country) and two summary questions \x{to report arousal-related and valence-related design features respectively} (\eg ``Please recall the data visualizations you have viewed and list the design factor(s) that helped increase the arousal level.''). Last, we interviewed the participants with a set of pre-prepared questions concerning the study design, such as ``Do you think the study instructed you well? Did you feel confused about anything?'', ``Do you think 20 is an appropriate number of images to view? Did you feel tired or impatient during the study?'', ``Do you think some images are significantly different from others in terms of eliciting your affects?''. 
On average, the participants spent about 20 minutes completing this study and 10 minutes on the interview.

\subsubsection{Results}

We collected both positive and negative feedback from the pilot study. For example, most participants agreed that 20 images is an appropriate amount for them to complete the study without losing patience. 
Also, the participants confirmed that the study task was clear and the whole process was well-guided.
However, on the downside, several participants reported that they did not feel confident in their ratings until they had viewed and rated several images (\eg ``\textit{Maybe there should be several examples in the very beginning to help me decide my criterion of rating.}''). Accordingly, we refined the introduction page by including three examples. The three examples were selected by a visualization expert to demonstrate the diversity of the stimuli.

\begin{figure}[t]
 \centering
 \includegraphics[width=\columnwidth]{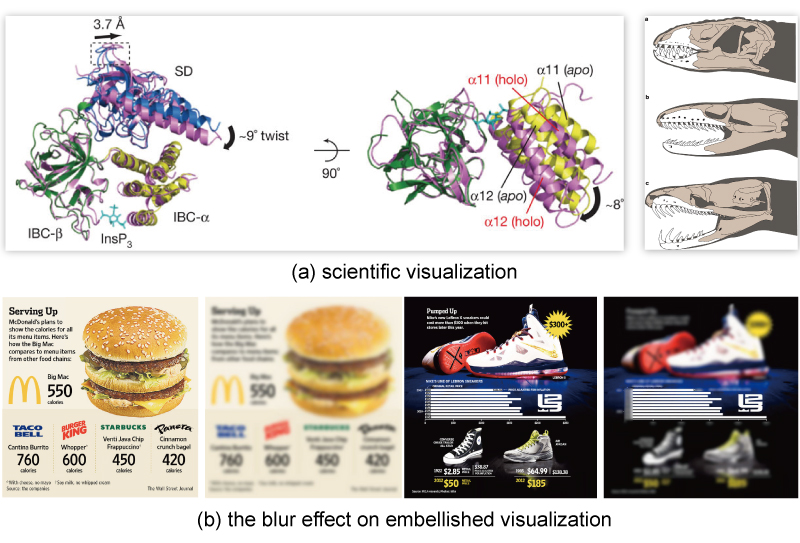}
  \vspace{-2em}
 \caption{\x{Scientific visualizations (a) and the comparison of the original version and the blurred version of two embellished visualizations (b).}}
 \label{fig:embellish}
 \vspace{-1em}
\end{figure}

We also noticed that there were some problems with our stimuli.
First, many participants reported that the scientific visualizations (see Fig.~\ref{fig:embellish}) were significantly different from others, since the participants were unfamiliar with such designs and felt these images should only be viewed by professional scientists or expert users (\eg ``\textit{This is not a visualization that I will come across in my daily life, and I feel it is apparently different from the other charts I saw.}'').
\x{Second, 
we found that embellishments are inherently rich in semantics and their semantics could hardly be erased by the blur effect. For example, a participant commented that "\textit{I can guess there is a McDonald hamburger through its shape and color. I love hamburgers!}", and another participant said "\textit{there is a gym shoe next to the chart and this aroused me.}" These two embellishments were arousing not because of their design, but because of what they represent (see Fig.~\ref{fig:embellish}). In other words, the existence of embellishments is likely to trigger a different mechanism of eliciting affective arousal and confound the effects of design with the effects of semantics.}
Third, several participants reported their discomfort with images that had a large aspect ratio. For example, one participant said, ``\textit{There was a lengthy chart, and I had to keep scrolling up and down to view it completely. This made it difficult to focus on my emotion.}''

Based on these findings, we removed scientific images, embellished images, and images with an aspect ratio greater than 3:1~\cite{borkin2013makes} from the \textit{targets410} corpus and kept 202 qualified images. However, a consequent problem was that the removal had changed the distribution of the image categories (\ie Infographic, Government, News Media), since many images removed because of embellishment belonged to the Infographic category. Thus, the sample sizes of the three categories became unbalanced. To deal with this problem, we searched the \textit{single2k} corpus to identify more qualified infographic images. Finally, we found 63 qualified images. After including these 63 images in our corpus (which resulted in 265 images in total), \x{the corpus contains 83 Infographic images (31\%), 82 News Media images (31\%), and 100 Government images (38\%)}. The distribution of the image categories is similar to that of the \textit{targets410} corpus, \x{thus ensuring the images are representative of the single visualization designs in the wild}. In our crowdsourcing study, we used this refined corpus as our stimuli.

\section{Crowdsourcing study}

Then, we conducted our formal study. We chose crowdsourcing to collect user data concerning the affective arousal elicited by data visualization design and explore design-arousal relationships.

\subsubsection{Methodology}

We recruited 184 participants from Prolific who speak English as their first language and have correct vision. We rejected the data from ten participants, as they misunderstood the study task (\eg failed to understand why the texts were blurred and kept complaining about the legibility) or provided data with poor quality (\eg giving all the images the same ratings without explaining why). The remaining 174 participants included 98 females, 75 males, and one identified as non-binary. Their ages ranged from 18 to 78 (\textit{M} = 38.47, \textit{SD} = 13.79), and their educational levels varied (Less than a high school diploma: 0.57\%, High school or equivalent: 31.03\%, Bachelor or equivalent: 48.28\%, Master or equivalent: 16.09\%, Doctoral or equivalent: 4.02\%).
The participants were paid \$9 per hour. 

The study procedure of the crowdsourcing study was identical to the pilot study. The participants first read an introduction page, then viewed 20 data visualizations (one at a time), rated arousal and valence, wrote down reasons for their ratings, and rated how they liked the designs. After viewing and rating all 20 images, the participants filled out a form to provide their demographic information and wrote down what design features they thought contributed to arousal and valence, respectively. 
The participants spent 22.34 minutes on average completing this study.

\subsection{\x{Analyses and Results}}

We got 3480 valid ratings for the 265 images, \x{including 1101 for the Infographic category, 1052 for the News Media category, and 1327 for the Government category}. Each image received 13.13 ratings on average.

\subsubsection{\x{Descriptive Analysis}}

\begin{figure}[b]
 \centering 
 \includegraphics[width=\columnwidth]{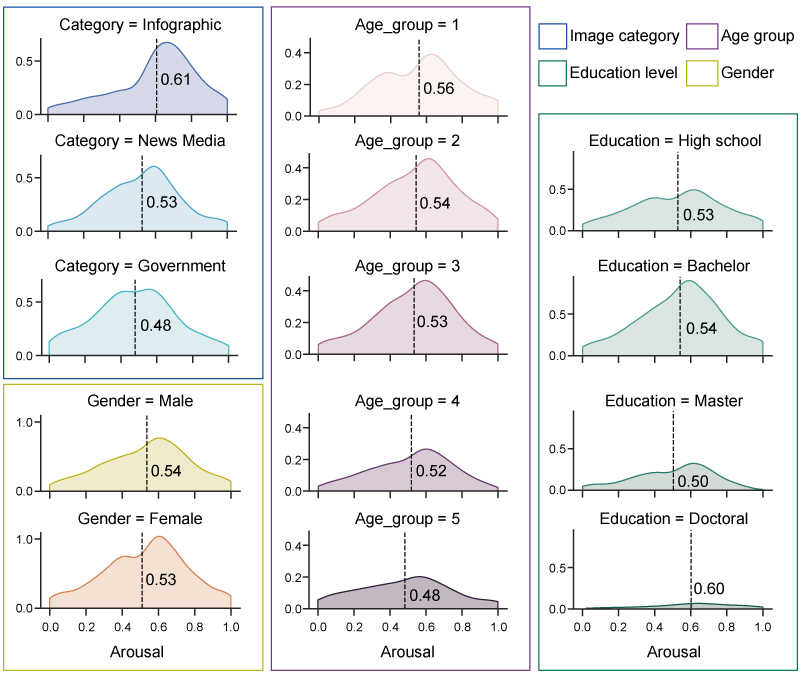}
 \caption{How the ratings of affective arousal distributed in different groups, including image categories, gender, age, and education level. Dashed lines represent means.}
 \label{fig:mean_comparison}
\end{figure}

\begin{figure*}[!t]
  \centering
  \includegraphics[width=\linewidth]{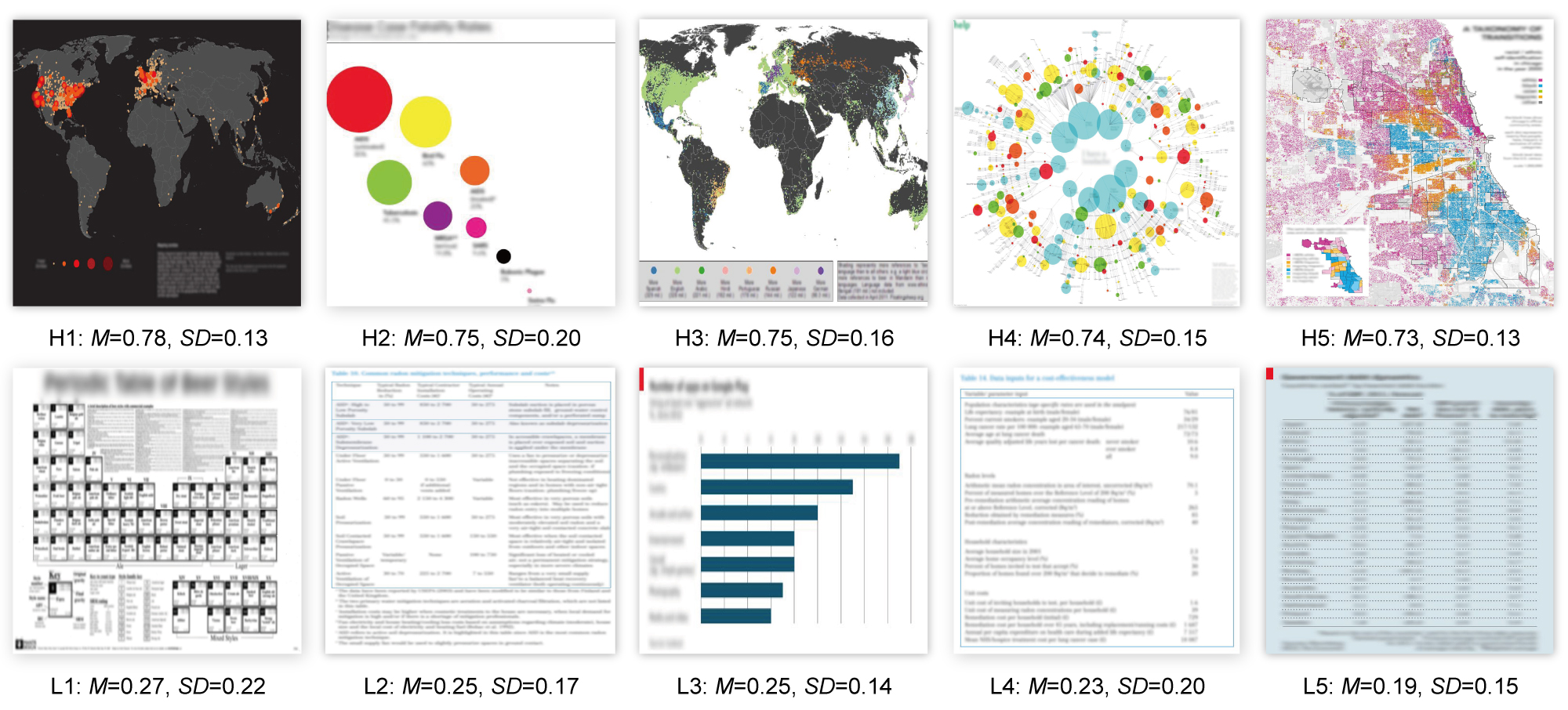}
  \caption{Five designs that received the highest scores in affective arousal (H1 to H5) and five designs that received the lowest scores (L1 to L5) in our study. The scores range from 0 to 1, where 0 means not aroused and 1 means strongly aroused. All texts were blurred.}
	\label{fig:teaser}
\end{figure*}

\paragraph{Consistency of Ratings}

\x{We examined whether people formed consistent judgments on the arousing level of the images.
We used intraclass correlation coefficients (ICC) to calculate the agreement in user ratings, and the result showed that the agreement was moderate (ICC1k = .589, 95\% CI = [.512, .658], \textit{p} < .001), suggesting that users held moderately similar judgments on affective arousal.}

\paragraph{Means \& Distributions Comparison}

Overall, the average arousal score of all the images was 0.53 (\textit{SD} = 0.23, 95\% CI = [0.52, 0.54]). \x{The distribution of the arousing levels of the images was close to a normal distribution}.
However, as shown in Fig.~\ref{fig:mean_comparison}, the means and distributions of affective arousal varied in different groups.
For example, the mean arousal of the \textit{Infographic} category was 0.61, higher than that of the \textit{News Media} category (\textit{M} = 0.53) and the \textit{Government} category (\textit{M} = 0.48). A Kruskal-Wallis H Test further showed that the difference was significant (\textit{H}(2) = 211.866, \textit{p} < .001). Dunn’s post-hoc test with a Bonferroni correction showed that the mean arousal of \textit{Infographic} was significantly higher than that of \textit{News Media} and \textit{Government} (\textit{p} < .001), and the mean arousal of \textit{News Media} was significantly higher than that of \textit{Government} (\textit{p} < .001).

We also examined the arousal scores set by different age, gender, and education groups. First, a Mann-Whitney U Test (\textit{z} = -.886, \textit{p} = .376) showed that there was no significant difference between the scores set by females and by males ("Non-binary" was not included because its sample size was too small). Then, we assigned the participants into 5 age groups (<=25, 26-35, 36-45, 46-55, >56) and compared the mean scores of arousal set by these groups. As shown in Fig.~\ref{fig:mean_comparison}, the mean arousal scores set by different age groups interestingly decreased with the increasing age. A Kruskal-Wallis H Test combined with a post-hoc test further showed that the difference was significant (\textit{H}(4) = 32.072, \textit{p} < .001) and mainly caused by the lower mean (\textit{M} = 0.48) of group 5 (age > 56). In other words, in our study, the younger participants were more aroused by the stimuli while the elder participants were generally calmer. A significant difference (\textit{H}(3) = 22.838, \textit{p} < .001) also existed in the scores given by participants with different education levels ("Less than a high school diploma" not included because its sample size was too small), mainly caused by the higher scores set by participants with a "Doctoral or equivalent" degree (\textit{M} = 0.60).
However, this finding should be taken cautiously since the sample size of "Doctoral or equivalent" was not big (7 participants).

We also calculated the mean arousal score for each image. Fig.~\ref{fig:teaser} shows the five images that received the highest scores in affective arousal and the five images that received the lowest scores. In line with the above findings, the top ranked images all belong to the \textit{Infographic} category (Fig.~\ref{fig:teaser} H1-H5) while the low-ranking images mostly belong to the \textit{Government} and \textit{News Media} categories (Fig.~\ref{fig:teaser} L2-L5).

\begin{figure}[t]
 \centering 
 \includegraphics[width=\columnwidth]{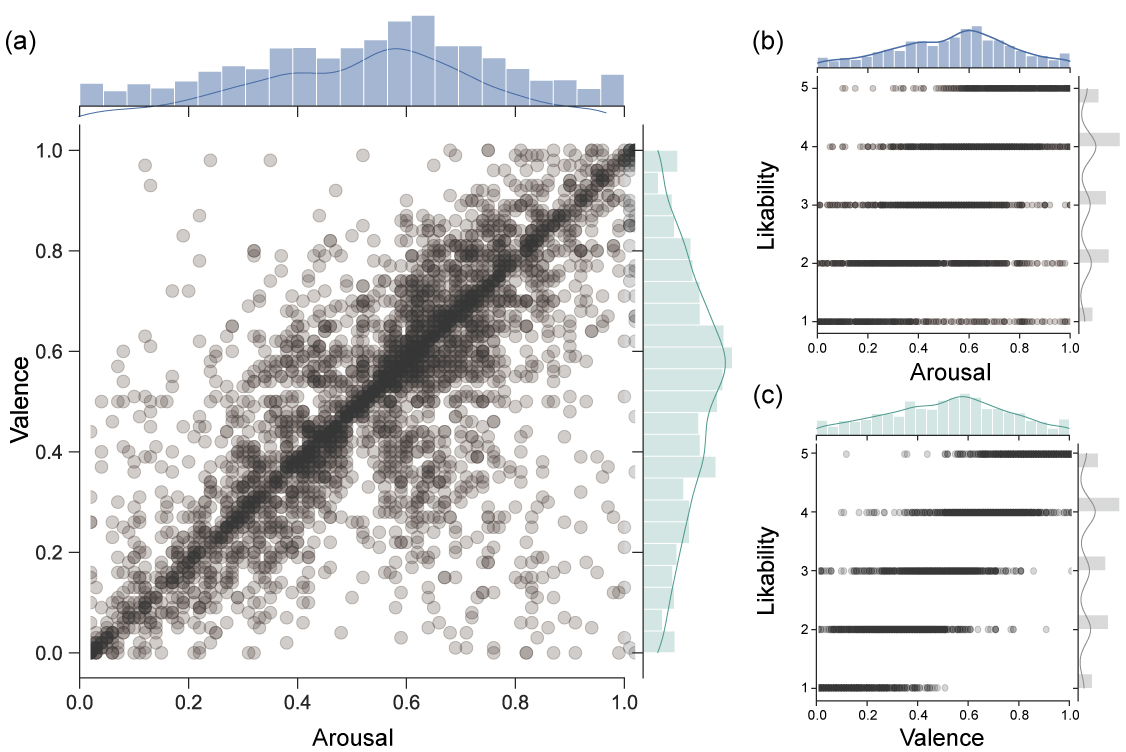}
 \caption{The relationships between (a) arousal and valence (\textit{$\tau$} = .592), (b) arousal and likability (\textit{$\tau$} = .539), and (c) valence and likability (\textit{$\tau$} = .736).}
 \label{fig:association}
\end{figure}

\paragraph{Correlation Analysis}

Then, we examined the correlations between affective arousal, valence, and likability \x{to see whether an arousing design will be pleasing or favorable}. Fig.~\ref{fig:association} plots the mutual relationships between these three variables. It can be seen from the figure that they generally showed a positive correlation with each other.
Given that likability is an ordinal variable and all the three variables are non-normal, we used Kendall's Tau (\textit{$\tau$}) to compute their correlation coefficients and significance levels. The results confirmed that arousal was positively correlated with likability (\textit{$\tau$} = .539, \textit{p} < .001), so as valence and likability \textit{($\tau$} = .736, \textit{p} < .001), whose positive correlation was even stronger. Besides, arousal and valence also showed a moderately positive correlation with each other (\textit{$\tau$} = .592, \textit{p} < .001). Note that these correlations still held when we conducted the correlational tests within different image categories.

To sum up, in general, high arousal was likely to co-occur with pleasure and high likability. However, there were also cases when arousal did not lead to pleasure or when arousal hindered likability. For example, as shown in Fig.~\ref{fig:association}, some participants set scores to  "high arousal + displeasure" or "high arousal + low likability". We will discuss such situations in detail in Section 5.2.

\begin{figure}[!b]
 \centering 
 \includegraphics[width=\columnwidth]{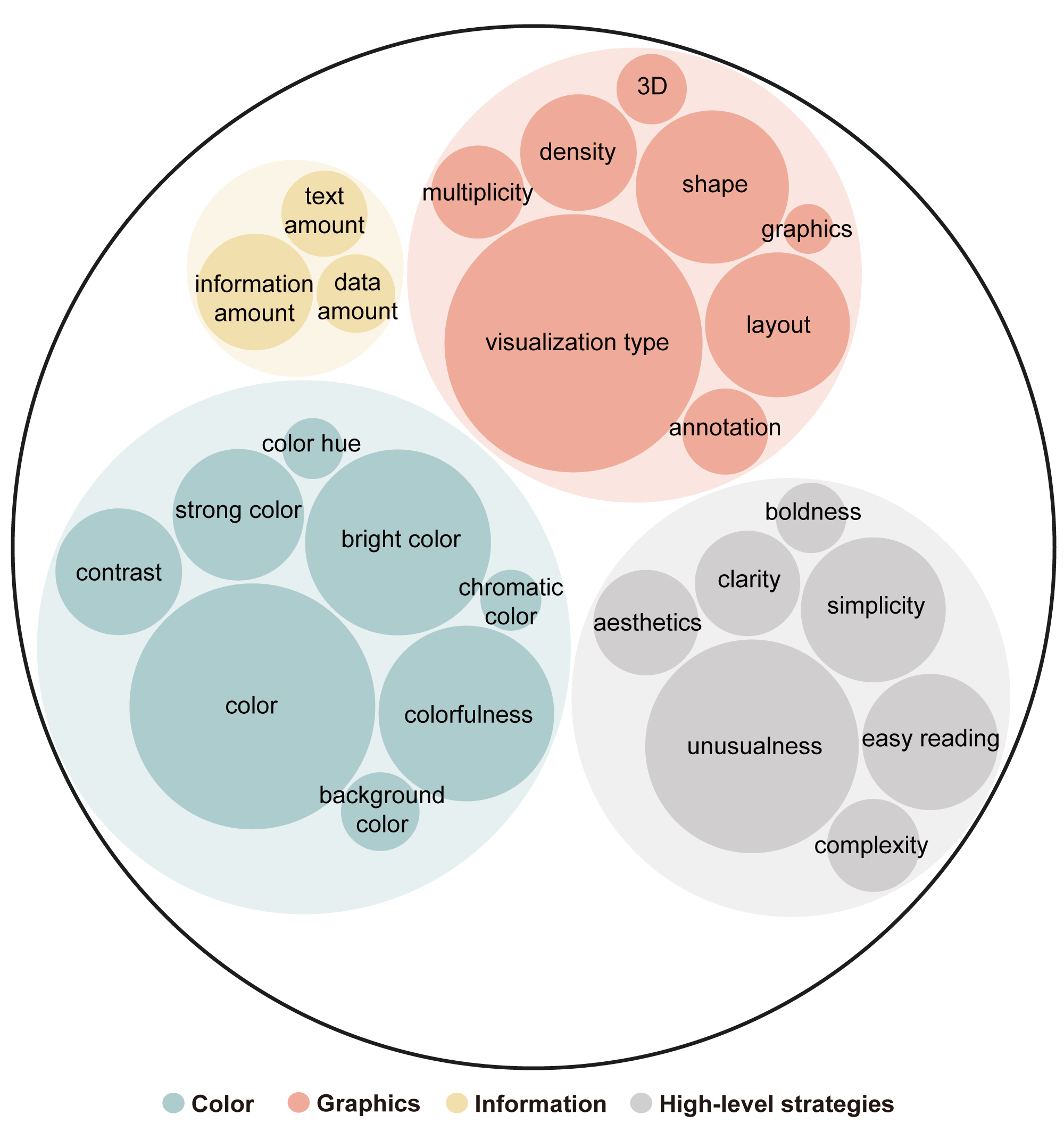}
 \caption{\x{Arousal-related features reported by the participants in our crowdsourcing study. The codes about color, graphics, and information are mapped to computable variables in Section 4.1.3.1. The high-level strategies are discussed in Section 5.1.}}
 \label{fig:themes}
\end{figure}

\subsubsection{Qualitative Analysis}

As stated above, at the end of the user study, we asked the participants to write down what design features had helped augment their affective arousal. This resulted in 174 comments that suggest arousal-related design features from the perspective of users.

Then, we conducted an in-depth qualitative analysis on these 174 comments following the bottom-up procedure of thematic analysis~\cite{braun2006using}. Two of the authors first went through all the user comments and took notes on the messages deemed useful. 
Then, they coded the comments independently and generated possible themes by grouping the codes.
After that, they met to compare codes and discuss mismatches.
Through iterative discussions, they generated three themes most commonly mentioned by the participants, including \textit{color}, \textit{graphics}, and \textit{information}.
Then, they assigned the low-level codes into these three themes and discussed whether to retain, modify, or merge the codes. 
For example, when describing colors that had strong visual impact, the participants used different expressions such as \textit{strong color}, \textit{vibrant color}, and \textit{bold color}. Therefore, these similar expressions were merged into one code, namely \textit{strong color}. 
It was also noticed that some codes described high-level design strategies rather than specific design features. For example, \textit{simplicity} can mean simple colors, simple visualization design, or a small amount of data. \textit{Unusualness} can refer to unusual color, unusual shape, or unusual visual encodings.
Therefore, such codes were categorized into another theme called \textit{high-level strategies}.
After iterative discussions, the coders achieved a 100\% agreement. At last, they generated 412 codes from the 174 comments in total. 
The frequencies of all the codes were analyzed and the result visualized, as in Fig.~\ref{fig:themes}. Note that for the clarity of presentation, codes that occurred only once were excluded. Collectively, the codes visualized in Fig.~\ref{fig:themes} constitute 92.72\% of all the codes. Thus, they are highly representative of all the arousal-related design features written down by the participants.

As shown in Fig.~\ref{fig:themes}, \textit{color} and \textit{graphics} were the most mentioned themes. Within the theme of color, 49 codes only broadly referred to \textit{color} as a contributor to affective arousal. Apart from this, \textit{bright color} (28), \textit{colorfulness} (25), \textit{strong color} (14), and \textit{contrast} (13) were the most frequently-mentioned specific color features.
Within the theme of graphics, \textit{visualization type} (54), \textit{layout} (17), \textit{shape} (16), and \textit{density} (11) were the most frequently-mentioned features.
Within the theme of information, 11 codes talked about \textit{information amount} in a broad sense, while six codes specifically referred to \textit{text amount} and five codes specifically referred to \textit{data amount}. 
Last, with the theme of high-level strategies, \textit{unusualness} (37), \textit{simplicity} (17), and \textit{easy reading} (15) were most mentioned. Besides, \textit{clarity} (9), \textit{aesthetics}(9), \textit{complexity} (5), and \textit{boldness}(4) were also repeatedly mentioned.


\subsubsection{\x{Inferential Analysis}}

\x{Then, we sought to predict the user ratings for affective arousal using these codes and identify the most significant design features.
For each of the codes about color, graphics, and information in Fig.~\ref{fig:themes}, we reviewed previous literature in relevant fields (\eg computer vision, multimedia) to find the mature and commonly-used methods to map these codes to computable variables.
Below we first introduce these features briefly, then report the modeling process.}

\paragraph{\x{Design Features}}
The mappings are summarized in Table.~\ref{tab:features}.


\begin{table*}[!t]
 \centering 
 \includegraphics[width=\textwidth]{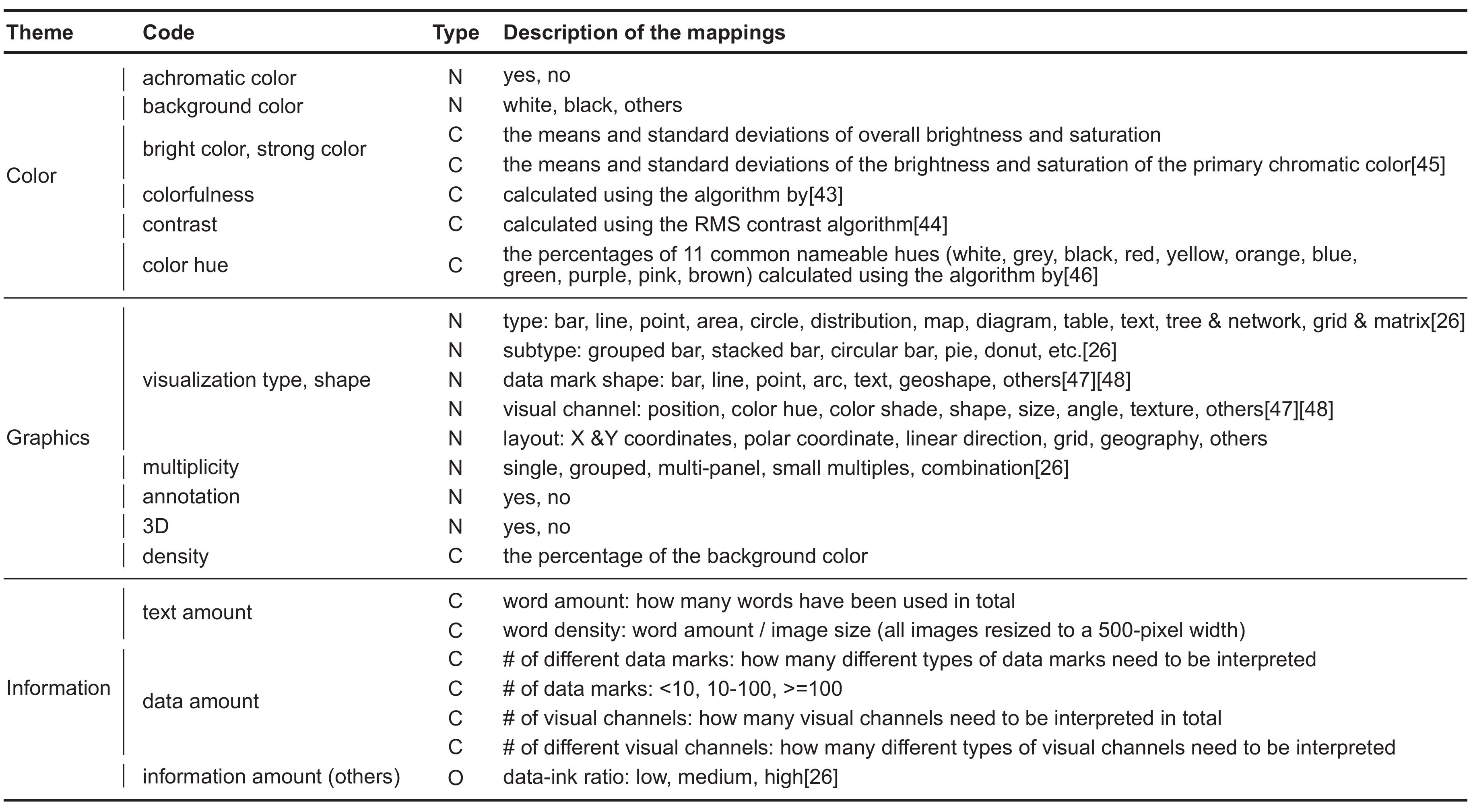}
 \caption{\x{The mappings from qualitative codes to variables. For variable types, N denotes nominal, C denotes continuous, and O denotes ordinal.}}
 \label{tab:features}
\vspace{-2em}
\end{table*}

\textbf{Color-based Features.}
\x{To translate the code \textit{chromatic color}, we used a nominal variable whose value is yes or no to depict whether a visualization design is achromatic. To translate the code \textit{background color} in Fig.~\ref{fig:themes}, we coded the images' background colors manually and found that white and black were the dominant choices. Therefore, this code was translated into a nominal variable whose value can be white, black, or others. For codes that can be measured as continuous variables, such as \textit{bright color}, \textit{strong color}, \textit{colorfulness}, and \textit{contrast}, we referred to prior literature~\cite{machajdik2010affective,reinecke2013predicting} and extracted corresponding design features, such as calculating the mean and standard deviation of the S (\ie saturation) and V (\ie brightness) channels in the HSV color space using OpenCV, computing colorfulness using the algorithm proposed by Hasler~\etal~\cite{hasler2003measuring}, and calculating contrast using the RMS contrast algorithm~\cite{peli1990contrast}.
Note that according to some comments, the participants' perception of brightness and saturation was also influenced by the dominant color, we also extracted the primary chromatic color of each image using Colorthief~\cite{colorthief} and calculated the mean and standard deviation of its saturation and brightness.}
Last, we translated the code \textit{color hue} using the algorithm proposed by van de Weijer~\etal~\cite{van2007learning}. This algorithm can identify and calculate the percentages of 11 nameable color hues (\ie black, blue, brown, green, gray, orange, pink, purple, red, white, yellow) in images. 

\textbf{Graphics-based Features.}
\x{
As two codes in this theme (\textit{visualization type}, \textit{shape}) were closely related to how data was visualized into graphical elements, we used five features to depict the visual encodings of the visualizations. 
First, following the typology in ~\cite{borkin2013makes}, data visualizations were categorized into 12 types (\eg \textit{bar}, \textit{area}, \textit{map}, \textit{circle}) and a set of subtypes (\eg \textit{grouped bar chart} and \textit{stacked bar chart} in the category of \textit{bar}; \textit{pie chart} and \textit{donut chart} in the category of \textit{circle}).
In addition, we used variables to characterize which data marks (\eg \textit{bar}, \textit{line}, \textit{geoshape}), visual channels (\eg \textit{position}, \textit{size}, \textit{shape}), and layout (\eg \textit{X and Y coordinates}, \textit{grid}, \textit{polar coordinate}) had been used according to existing taxonomies~\cite{satyanarayan2016vega,munzner2014visualization}.
For example, H4 in Fig.~\ref{fig:teaser} was characterized as a visualization type called \textit{area} and a subtype called \textit{proportional area chart}. The shape of its data marks is \textit{circle}, and the overall shape of this visualization is also circular because it organizes the marks using a \textit{polar coordinate}. Two visual channels (\textit{color} and \textit{position}) have been used to encode data. 
For the remaining qualitative codes in this theme, we mapped \textit{multiplicity}, which defines whether the visualization is stand-alone or somehow grouped with other visualizations, to categories such as \textit{single}, \textit{grouped}, and \textit{multi-panel}~\cite{borkin2013makes}.
We mapped \textit{annotation} and \textit{3D} to nominal variables ([\textit{yes}, \textit{no}]) and computed \textit{density} by calculating the white space (the percentage of the background color) of each image.}

\textbf{Information-based Features.}
Information-based features concern the cognitive load of viewing a data visualization and can be further split into two main types: texts and data.
First, although all the texts had been blurred, \x{several participants stated that they could still “feel” that there were a lot of texts to be read and this had also impacted their arousal.}
Therefore, we used Tesseract-OCR~\cite{smith2007overview} to identify the texts on the images, manually revised inaccurate identifications, and then counted how many words each image contained and the word density on each image by dividing the word amount by the size of the image. 
Next, to quantify the amount of data, we counted how many visual channels were used in each image in total, as well as how many distinct visual channels were used. We also counted the number of data marks in each image and how many distinct data marks were used. For example, Fig.~\ref{fig:teaser} H2 only has eight data marks while H4 contains more than 100 data marks, but both of them only used one type of data mark, namely \textit{circle}. 
In some images, the data marks were too many to be counted precisely (\eg thousands of overlapping dots), therefore, we transformed this feature into a nominal variable with three categories (\textit{<10}, \textit{10-100}, \textit{>=100}).
\x{Last, we used an ordinal variable, data-ink ratio ([\textit{good}, \textit{medium}, \textit{bad}])~\cite{borkin2013makes}, to characterize the cognitive load brought by information that was neither texts nor data (\eg axis ticks, gridlines).}


\begin{figure*}[!t]
 \centering
 \includegraphics[width=\textwidth]{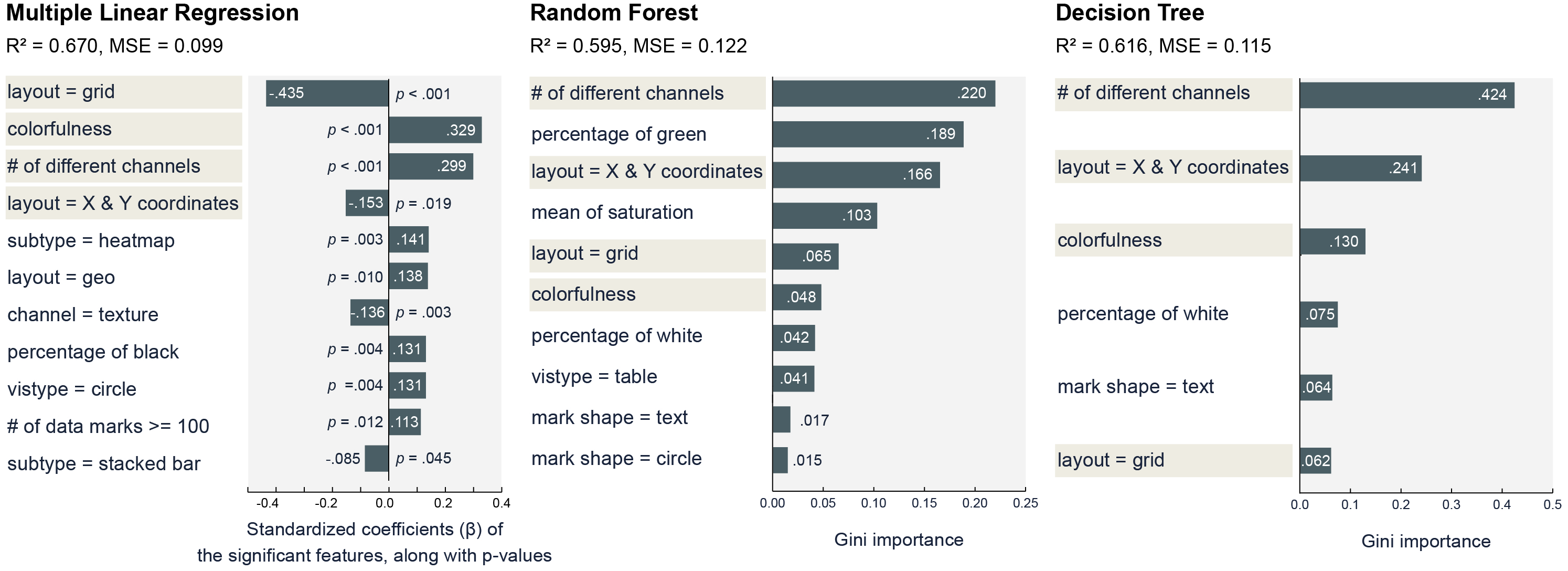}
 \caption{\x{Comparing the important features suggested by different models. Note that the Multiple Linear Regression Model indicates importance through statistical significance (p-values) and standardized coefficients, while Random Forest and Decision Tree suggest key features through Gini importance. ${R}^2$ and MSE show the model performance, where ${R}^2$ (the coefficient of determination) represents the proportion of the variance explained, and MSE (mean squared error) measures the difference between the original and predicted values.}}
 \label{fig:features_compare}
 \vspace{-1em}
\end{figure*}

\paragraph{Modeling affective arousal}
\x{Based on the extracted features, we tried five widely used predictive models to explore important moderators of affective arousal. Three are linear models (Multiple Linear Regression, Ridge Regression, Lasso Regression) and two are ensemble models (Random Forest, Decision Tree). These models can help us estimate the relationship between affective arousal and design features while interpreting which features are contributing to the relationship.}
The dependent variable is the mean arousal scores of the 265 images, and the design features of the 265 images we extracted in Table.~\ref{tab:features} are independent variables.
During the data pre-processing stage, first, we checked and revised empty and mistaken values.
Then, for all nominal variables, we used the one-hot encoder in SciPy to transform them into a set of binary variables that use 1 and 0 to indicate the presence or absence of a category~\cite{garcia2015data}.
Next, we tried re-scaling the raw arousal scores set by each participant before calculating the mean arousal score received by each image. This step was done because we noticed from both the pilot study and the crowdsourcing study that the participants showed varied rating habits. For example, even with a very similar judgment of a design, some participants tended to set higher scores to arousal (\eg 0.95) while some thought 0.70 or 0.80 were already high scores. In other words, the participants' criteria of rating affective arousal were differently-ranged. Therefore, re-scaling the ratings helped mitigate the influence of rating habits and revealed the real judgments of all the participants. Specifically, we tried two common means of re-scaling, including the z-score normalization and the Min-Max normalization~\cite{garcia2015data}. 
When constructing the models, we tested the performance of the two re-scaled dependent variables as well the un-scaled version. Results showed that the z-score normalization method performed the best. Therefore, we adopted this version of dependent variable when constructing the final models.

After data pre-processing, we used Pearson correlation coefficients to get an initial understanding of the correlations between the independent variables and the dependent variable.
Results showed that features such as the \textit{number of different visual channels} (\textit{r} = 0.52, \textit{p} < .001), \textit{colorfulness} (\textit{r} = 0.48, \textit{p} < .001), and \textit{standard deviation of saturation} (\textit{r} = 0.46, \textit{p} < .001) had a significantly positive relationship with affective arousal, while features such as \textit{X-Y coordinates layout} (\textit{r} = -0.61, \textit{p} < .001), \textit{table visualization} (\textit{r} = -0.52, \textit{p} < .001), and \textit{percentage of white} (\textit{r} = -0.49, \textit{p} < .001) had a significantly negative relationship with affective arousal. High correlations also exist between some design features. For example, the \textit{percentage of white} was highly correlated with \textit{mean brightness} (\textit{r} = 0.82, \textit{p} < .001). \x{These correlations were referred to when performing feature selection to prevent the overfitting problem.}

\x{Next, we fed the variables into the five models one by one. To enhance the validity of results, we adopted the idea of machine learning and split the dataset into a training set (70\%) and a test set (30\%) using Sklearn. When splitting the dataset, we also guaranteed that the distribution of the values of arousal in the test set was similar to that of the training test, which was close to a normal distribution. Then, we fed the training set to the models and evaluated the goodness of fit using the test set. }

\x{\textbf{Linear Models.}
For the Multiple Linear Regression, we fine-tuned the models by (1) fitting the models using a stepwise approach to drop redundant features;} (2) conducting multicollinearity diagnostics using the Variance Inflation Factor (VIF). A VIF above 10 implies serious multicollinearity and a VIF above 5 is usually deemed problematic~\cite{james2013introduction};
(3) Ensuring that there was no autocorrelation problem using the Durbin-Watson test (0 < DW < 2); (4) Ensuring that the residuals were normally distributed by conducting the Anderson-Darling test and plotting a histogram for residuals. 
\x{For the Lasso Regression and Ridge Regression, we used five-fold cross-validation to identify the optimal values of their key parameter, alpha (\ie the strength of regularization). As a result, we found that in our case, the optimal alpha values for these two models were 0, which made their results identical to that of the Multiple Linear Regression.} 

\x{As shown in Fig.~\ref{fig:features_compare}, the Multiple Linear Regression yielded 11 statistically significant features. Seven features showed a significantly positive correlation with affective arousal, including \textit{colorfulness} ($\beta$ = .329, \textit{t} = 7.536, \textit{p} < .001), \textit{the number of different visual channels} ($\beta$ = .299, \textit{t} = 5.828, \textit{p} < .001), \textit{visualization subtype = heatmap} ($\beta$ = .141, \textit{t} = 3.013, \textit{p} = .003), \textit{layout = geo} ($\beta$ = .138, \textit{t} = 2.588, \textit{p} = .010), \textit{percentage of black} ($\beta$ = .131, \textit{t} = 2.957, \textit{p} = .004), \textit{visualization type = circle} ($\beta$ = .131, \textit{t} = 2.886, \textit{p} = .004), and \textit{number of data marks >= 100} ($\beta$ = .113, \textit{t} = 2.531, \textit{p} = .012).
On the other hand, four features showed a significantly negative correlation with affective arousal, including \textit{layout = grid} ($\beta$ = -.435, \textit{t} = -7.279, \textit{p} < .001), \textit{layout = X and Y coordinates} ($\beta$ = -.153, \textit{t} = -2.360, \textit{p} = .019), \textit{channel = texture} ($\beta$ = -.136, \textit{t} = -3.034, \textit{p} = .003), and \textit{visualization subtype = stacked bar} ($\beta$ = -.085, \textit{t} = -2.020, \textit{p} = .045). The training set explained 72.6\% of the variance in the dependent variable (${R}^2$ = .726, MSE = .071), and the test set explained 67.0\% (${R}^2$ = .670, MSE = .099).}

\x{\textbf{Ensemble Models.}
For ensemble models, we first identified the optimal hyperparameters (\eg max depth) for the models by performing five-fold cross-validation within a grid of hyperparameter ranges. We also fine-tuned the hyperparameters by observing the performance of the training set and the test set to avoid overfitting. 
At last, the Random Forest model had a max depth of 2. After iterating 1000 times, the average ${R}^2$ of the training set was .644 (MSE = .092), and the average ${R}^2$ of the test set was .595 (MSE = .122). 
The Decision Tree model had a max depth of 3. The average ${R}^2$ of the training set was .668 (MSE = .085), and the ${R}^2$ of the test set was .616 (MSE = .115). We also computed the Gini importance of the features in these two models and visualized high-ranking features in Fig.~\ref{fig:features_compare}. The most important features in the Random Forest model included the \textit{number of different visual channels} (Gini importance = .220), \textit{percentage of green} (.189), \textit{layout = X and Y coordinates} (.166), \textit{mean of saturation} (.103), \textit{layout = grid} (.065), \textit{colorfulness} (.048), \textit{percentage of white} (.042), \textit{visualization type = table} (.041), \textit{mark shape = text} (.017), and \textit{mark shape = circle} (.015). These 10 features collectively explained 90\% of the total importance. The most important features in the Decision Tree model included the \textit{number of different visual channels} (.424), \textit{layout = X and Y coordinates} (.241), \textit{colorfulness} (.130), \textit{percentage of white} (.075), \textit{mark shape = text} (.064), and \textit{layout = grid} (.062). These six features collectively explained 99\% of the total importance. 
}

\x{\textbf{Summary.}
By comparing the results in Fig.~\ref{fig:features_compare}, we found that four design features were collectively supported by all the models as the key moderators of affective arousal, including the \textit{number of different visual channels}, \textit{colorfulness}, \textit{layout = X and Y coordinates}, and \textit{layout = grid}. The first two features showed positive effects on affective arousal in all models, while the latter two features consistently showed a negative relationship with affective arousal. In other words, in our exploratory study, we found that the participants were more aroused by colorful visualizations with a certain amount of visual complexity. However, conventional layouts that put data in an upright and square space were less arousing.
We also noticed that there were features whose importance was not cross-validated by different models. For example, the \textit{percentage of green} showed high importance in the Random Forest model, but it was not important in other models. Such features are worthy of more investigation in the future.}

\section{Discussion}

Below we discuss the design implications that arise from this work, the possible negative consequences of eliciting high arousal, \x{the challenges of automatic visualization design evaluation, as well as limitations and future work.}

\subsection{Design Implications}

\ul{First, we contribute new knowledge for understanding how data encodings influence affective arousal.} An interesting finding of this work is that in data visualization design, graphical elements that are related to data encodings showed significant capability of eliciting affective arousal. 
For the visualization community, this finding is significant in two main ways.
First, we have found new evidence that the appearance of data visualization does affect user experience. In the past, Moere~\etal~\cite{moere2012evaluating} found that when encoded with identical data, different chart types (\eg treemaps and sunburst diagrams) triggered different subjective feelings and preferences.
Borkin~\etal~\cite{borkin2013makes} found that people remembered certain chart types better than others. In other words, choosing data encodings is not only about how to present data but also about manipulating user experience. 
Second, this work also supplements the existing knowledge about user-centered visualization design from the perspective of affective arousal. For example, in the study by Borkin~\etal~\cite{borkin2013makes}, the most remembered visualization types were \textit{grid \& matrix} and \textit{tree \& network}.
\x{However, in our study, visualizations that use a grid layout (\eg L1 in Fig.~\ref{fig:teaser}) were less arousing than others.}
This suggests that viewing visualization design from the perspective of affective arousal may be different from viewing it from perspectives such as memorability and aesthetics (although these perspectives are all user-centered). In other words, focusing on different aspects of user experience may also lead to different focuses of design.

\x{\ul{Second, we found colorfulness outweighs other color-based features in terms of moderating affective arousal in the context of data visualization design.} 
Although the affective power of color has long been verified in various domains, in different research contexts, the findings can be different.
For example, red was identified as the most arousing color in classic psychological experiments~\cite{jacobs1974effects}, but some researchers also found evidence that green can be more arousing than red~\cite{valdez1994effects}.
Our work, however, found that colorfulness is more related to arousal than specific color hues in visualization design. According to our study, a reason may be that people's color perception would be influenced by their attempts to decode data when viewing a data visualization.
For example, some participants reported low arousal to designs with a large amount of red because the color had hindered reading (\eg "\textit{it is too bright and dazzling}"). Similarly, some participants stated that the use of analogous color hues (\eg red and orange) had made the graphical elements on the charts indistinguishable, thus lowering arousal. On the other hand, colorfulness can be both viscerally stimulating and useful in presenting data, thus making it the most arousing color-based feature compatible with data visualization design.
}

\ul{Third, data visualizations that belonged to the \textit{Infographic} category elicited stronger affective arousal than others.}
Infographics are often seen as a visual form that pays more attention to data communication and appealing to viewers subjectively than traditional visualizations~\cite{lan2021smile,lankow2012infographics}, and previous work~\cite{wang2018infonice,byrne2015acquired} has often focused on studying the rich embellishment techniques in infographics such as pictograms, illustrations, and visual metaphors. 
However, another interesting finding from this work is that even after we removed all embellished designs for our corpus, the images in the \textit{Infographic} category were still significantly more arousing than others. This finding indicates that the specificity of infographic design may not only lie in embellishment, but also in how it deals with color, graphics, and information. As shown by an empirical study~\cite{bigelow2014reflections}, when creating infographics, designers often spend a lot of time choosing encoding approaches and polishing the appearance of the visualization iteratively, because the designers not only hope to present data clearly, but also to create something new and unique. Such needs meet the \textit{unusualness} strategy we mentioned above and may help improve the arousing level of a design.

\x{\ul{Fourth, the qualitative and quantitative analyses collectively imply the strategies for manipulating arousal.} 
In the crowdsourcing study, we identified several high-level strategies for eliciting affective arousal by coding user comments qualitatively (Fig.~\ref{fig:themes}). After constructing models to predict arousal, we found that the quantitative results suggested by the models resonate more or less with the qualitative codes.
For example, our models showed that affective arousal is significantly higher when a visualization does not use layouts such as Cartesian coordinates, and this resonates well with the most mentioned high-level strategy by the participants, namely \textit{unusualness} (\eg "\textit{The design factors that helped increase the arousal was those that looked interesting and new, something I've never seen before.}"). Another two strategies mentioned by the participants (\textit{boldness} and \textit{aesthetics}) were partly supported by the significance of colorfulness, as previous research has found colorfulness is linked with the perception of energization and extraversion~\cite{pazda2019color}, as well as an indicator of aesthetics~\cite{harrison2015infographic,reinecke2013predicting}.}
However, there seems to be controversy over \textit{simplicity} and \textit{complexity} (we can see the co-existence of these two strategies in Fig.~\ref{fig:themes}). Our models, however, ended up leaning towards the side of \textit{complexity} because affective arousal showed a significantly positive relationship with the \textit{number of different visual channels}. \x{However, by referring to the raw comments written by the participants, we found that \textit{simplicity} and \textit{complexity} may not necessarily contradict each other. 
For example, when reporting that simple designs had increased their affective arousal, some participants were in fact emphasizing the understandability of the designs (\eg ``\textit{This diagram looks like a really simple way to display data, even to a layperson}''). On the other hand, when reporting that complex designs were more arousing, some participants were emphasizing the richness of visuals (\eg "\textit{different graphs and charts - when there was more information on the chart to concentrate on the arousal was higher}"). In other words, for designers, it may be beneficial to include an appropriate level of visual complexity while ensuring that the data visualization is understandable.}



\subsection{The Negative Side of High Arousal}

As we have discussed in Section 4.1.1.3, in our exploratory study, high affective arousal did not always lead to high pleasure or high preference. In some cases, the participants did not like or enjoy data visualization designs that elicited high arousal, or they thought high arousal was harmful. By analyzing such comments, we identified two main reasons:

\ul{First, high arousal may lead to physical discomfort.}
In our crowdsourcing study, several participants reported high arousal but thought the corresponding designs were too chaotic and over-stimulating (\eg ``\textit{It is a very arousing visual in terms of color and design, however looking at it makes your eyes go unwell so does not give great pleasure when looking at the visual.}'', "\textit{Unpleasant over-stimulating colours, chaotic graphic that is difficult to decipher and is unpleasant to look at.}"). One participant commented that "\textit{Too much red is making me feel nervous.}" 
These findings resonate with prior psychological studies that intense affects may bring about side effects~\cite{hyman1990ethics} or even evolve into tension and anxiety~\cite{malmo1957anxiety}. Therefore, for designers, designing an arousing data visualization is not equal to simply adding up all stimulating design features. On the contrary, it is necessary to find a balance between arousing affects and preventing discomfort.

\ul{Second, high arousal may be caused by inappropriate design.}
In some cases, the participants were affectively aroused because the designs looked confusing or ugly to them. For example, one participant said that a design "\textit{looks really cool but I don't know how easy it would be to interpret data using this}". Similarly, another participant wrote: "\textit{A very striking chart that looks like it makes sense only to the person who created it.}" We also saw comments that reported high arousal accompanied by negative affects such as anger and irritation (\eg "\textit{I don't like the orange. It makes me angry.}", "\textit{The red line is irritating.}"). Under such circumstances, high arousal is not beneficial.

\subsection{Towards Automatic Design Evaluation}

Understanding what design features contribute to affective arousal has paved a path toward automatically evaluating the arousing level of data visualizations or automatically generating arousing data visualizations. For example, we can envision a feedback system that can intelligently assess the arousing level of a data visualization design by scoring its design from multiple arousal-related aspects and identifying its strengths and weaknesses, thus guiding designers to improve their work. 
\x{To achieve such goals, extracting computable features from data visualization designs automatically becomes a pursuit. In this work, although about half of the features were calculated by computer, we noticed that there are three main challenges if we want to fully automate the process of feature extraction.}

\x{\ul{First, not all the design-related concepts can be easily transformed into computable features.} For example, so far, there is no recognized algorithm for evaluating the data-ink ratio of a data visualization automatically. Besides, high-level concepts such as \textit{unusualness} are difficult to be measured computationally, as they are usually feelings co-shaped by multiple design elements and are moderated by the participants' own knowledge, values, or preferences. How to take advantage of the valuable signals that arise from such expressions is a tricky problem.}

\x{\ul{Second, parsing the composition of a data visualization from an image can be difficult.} Understanding how a data visualization is composed of data, graphical elements, and texts requires utilizing techniques such as object detection, recognition, and classification. When the input is pixel-based images, the accuracy of the outcomes may not be satisfying.
For example, when calculating the whitespace, the algorithm sometimes misunderstood which color is the background color. When detecting texts from the images, the algorithm frequently made mistakes. As a result, we had to adjust these results manually. 
Therefore, to achieve automatic design evaluation, we still need to incorporate more state-of-the-art techniques from computer vision and develop more models and algorithms customized for data visualization design to improve the accuracy of design feature extraction.
}

\x{\ul{Third, how to assess the quality of a design remains a challenge.} Currently, we can use variables to describe the existence of a design feature or to what extent it is used. However, we are not able to judge whether the feature is used "appropriately". For example, in our study, some participants reported low arousal to visualizations that were badly designed (\eg “\textit{Not much immediate clarity what the visualisation is about hence my interest is low and disinterested.}”).
In other words, except for "quantifying" the design, the "quality" of a design also counts; except for "describing" the design, "assessing" the design also counts. This has posed another challenge for future research.}


\subsection{Limitations and Future Work}

\x{This work is exploratory in nature so the findings of this work were mainly limited to correlations rather than causal effects and were constrained by the inherent limitations of exploratory studies.}
First, although we have carefully constructed a corpus that contains diverse data visualization designs, the corpus is still by no means exhaustive.
Second, affective arousal is a subjective feeling that is difficult to quantify. To mitigate this problem, we excluded the influence of semantics, sampled diverse participants from a large population pool through crowdsourcing, and normalized each participant's ratings when constructing the models. However, noise caused by subjectivity may have inevitably entered and lowered the accuracy of prediction. 
Third, the features we extracted in this work are not exhaustive and are limited in characterizing some high-level subjective concepts such as aesthetics and unusualness. Also, \x{this work is short in revealing the fundamental impact of low-granularity design elements (\eg angles, roundness) on affective arousal. To address such issues, more controlled experiments should be done to examine the effects of a certain design condition while other variables are rigorously controlled.}
\x{We also hope a follow-up study can be done in the future to test the generalizability of this study by collecting more samples and systematically manipulating one or more of the identified features.}


We see several directions for future research.
First, 
\x{more alternative methods can be utilized to deepen our understanding of the design-affect relationship in data visualization.
For example, by excluding any semantics, this work did not examine the interaction between content and visualization design. Future work could fill this gap by first measuring the arousal caused by content and then analyzing how much arousal is augmented or lessened by design.}
\x{In addition, future work could examine how the incorporation of embellishment moderates the conveyance of semantics and the elicitation of affects.}
\x{Second, as our study has suggested that affective arousal may be related to the complexity of data visualization, future work could further investigate how people's visualization literacy influences their perceived complexity and arousal.}
\x{Third, more work can be done to investigate another important dimension of affects, namely valence, such as how valence-related design features are different from arousal-related features and how they interact with each other.}
Last, we hope there will be more research into how we should take advantage of the benefits brought by affective arousal and how we should avoid its negative side effects.
\x{This asks for the incorporation of more empirical knowledge, such as understanding the practice of designers, to help outline clearer guidelines for what should and should not be done.}
\section{conclusion}

\x{This work explores the relationship between visualization design and affective arousal.}
First, we collected a corpus of 265 data visualizations and conducted a crowdsourcing study with 184 participants to gather ratings on the affective arousal elicited by their design \x{as well as qualitative feedback}.
\x{Based on the study data, we first manually coded the arousal-related design features reported by the participants, then mapped these features to computable variables and constructed regression models to identify the most significant features.
By comparing the results of the models, we finally identified four design features (\eg colorfulness, the number of different visual channels) that were cross-validated as important features correlated with affective arousal.
We hope this work will inspire more future research on affective visualization design and user experience with data visualization.
}



\ifCLASSOPTIONcaptionsoff
  \newpage
\fi



%

\bibliographystyle{IEEEtran}
\bibliography{main}




%

\begin{IEEEbiography}
    [{\includegraphics[width=1in,height=1.25in,clip,keepaspectratio]{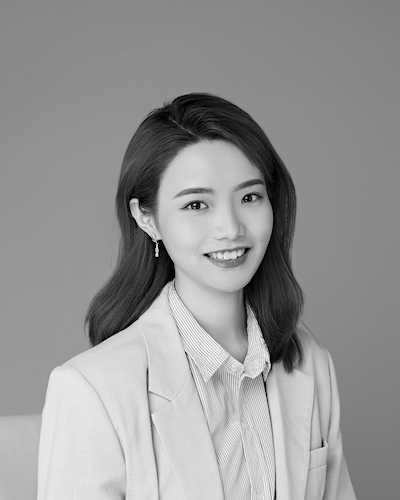}}]{Xingyu Lan} received her Ph.D. degree in design from Tongji University, China, in 2022. 
    She is going to work as an assistant professor at the College of Journalism, Fudan University, China. She conducts interdisciplinary research across fields such as data visualization, data journalism, digital communication, and design. Her research interests include data-driven storytelling, visualization design, human-computer interaction, and visualization for the masses.
\end{IEEEbiography}
\vskip -2\baselineskip plus -1fil
\begin{IEEEbiography}
    [{\includegraphics[width=1in,height=1.25in,clip,keepaspectratio]{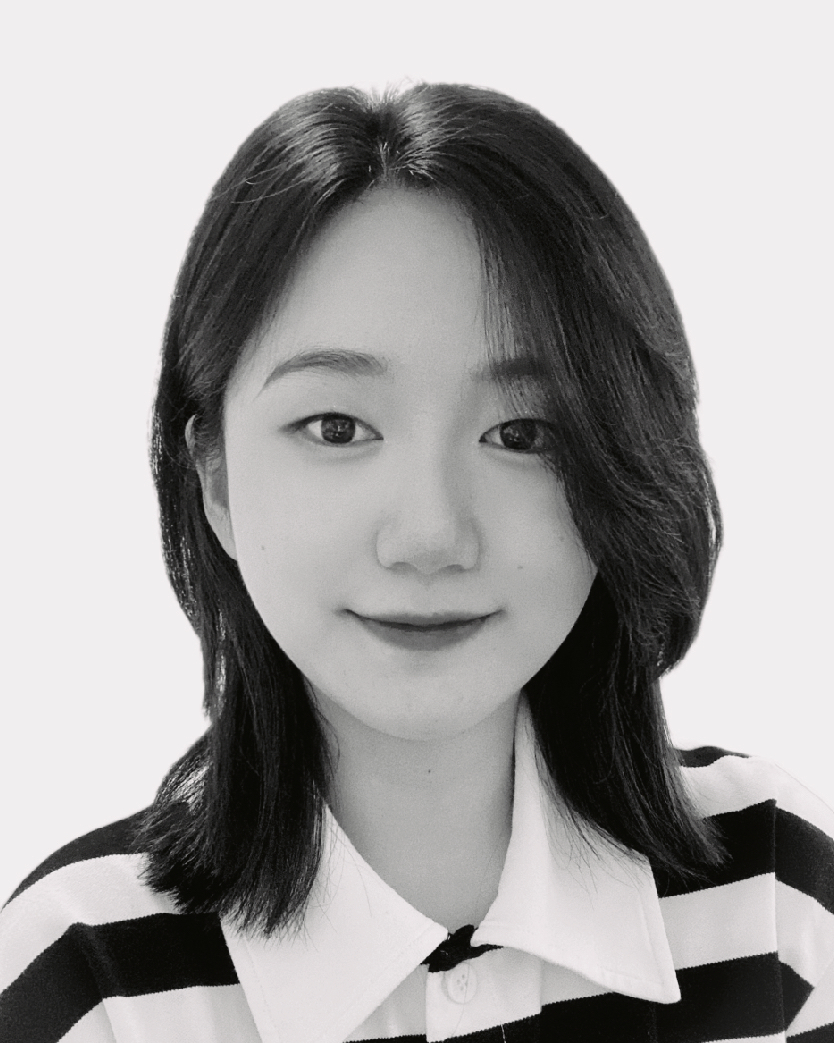}}]{Yanqiu Wu} received her MA degree in design from Tongji University, China, in 2022. She is currently a research assistant with Intelligent Big Data Visualization Lab (iDV$^x$ Lab), the College of Design and Innovation, Tongji University, Shanghai, China. Her research interests include data visualization design, intelligent user interfaces, and human-computer interaction.
\end{IEEEbiography}
\vskip -2\baselineskip plus -1fil
\begin{IEEEbiography}
    [{\includegraphics[width=1in,height=1.25in,clip,keepaspectratio]{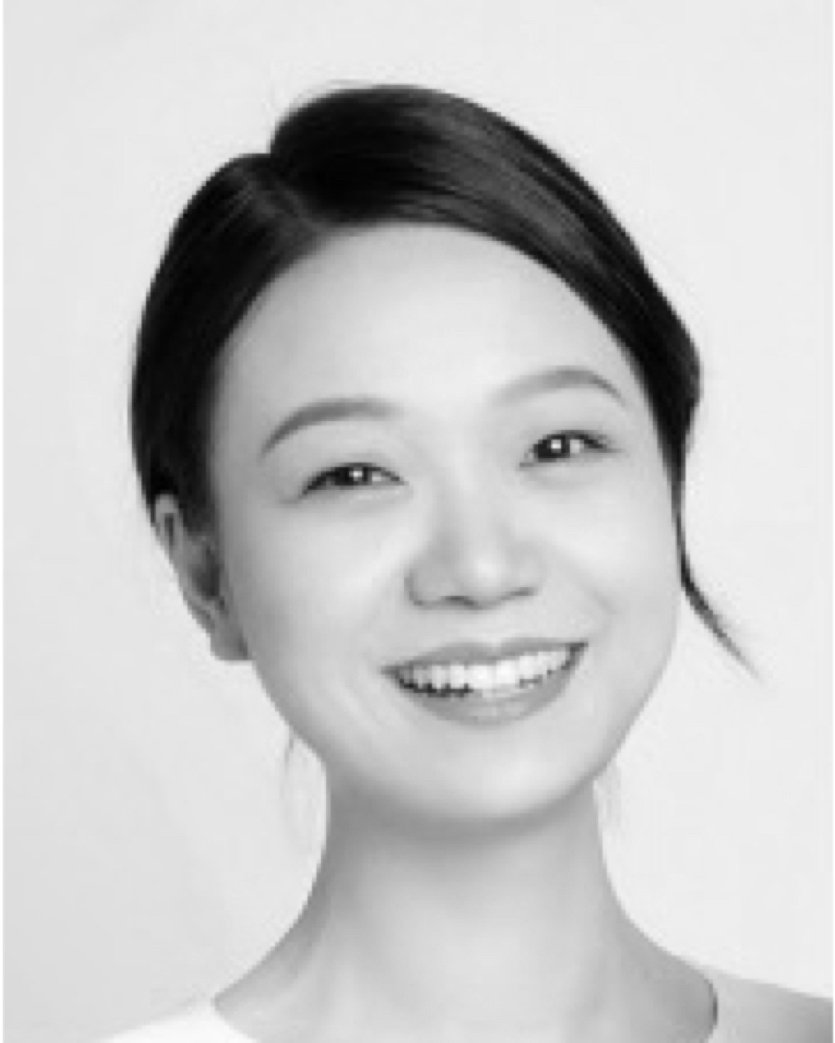}}]{Qing Chen} received her B.Eng degree from the Department of Computer Science, Zhejiang University and her Ph.D. degree from the Department of Computer Science and Engineering, Hong Kong University of Science and Technology (HKUST). After receiving her PhD degree, she worked as a postdoc at Inria and Ecole Polytechnique. She is currently an assistant professor at Tongji University. Her research interests include information visualization, visual analytics, humancomputer interaction, online education, visual storytelling, intelligent healthcare and design.
\end{IEEEbiography}
\vskip -2\baselineskip plus -1fil
\begin{IEEEbiography}
    [{\includegraphics[width=1in,height=1.25in,clip,keepaspectratio]{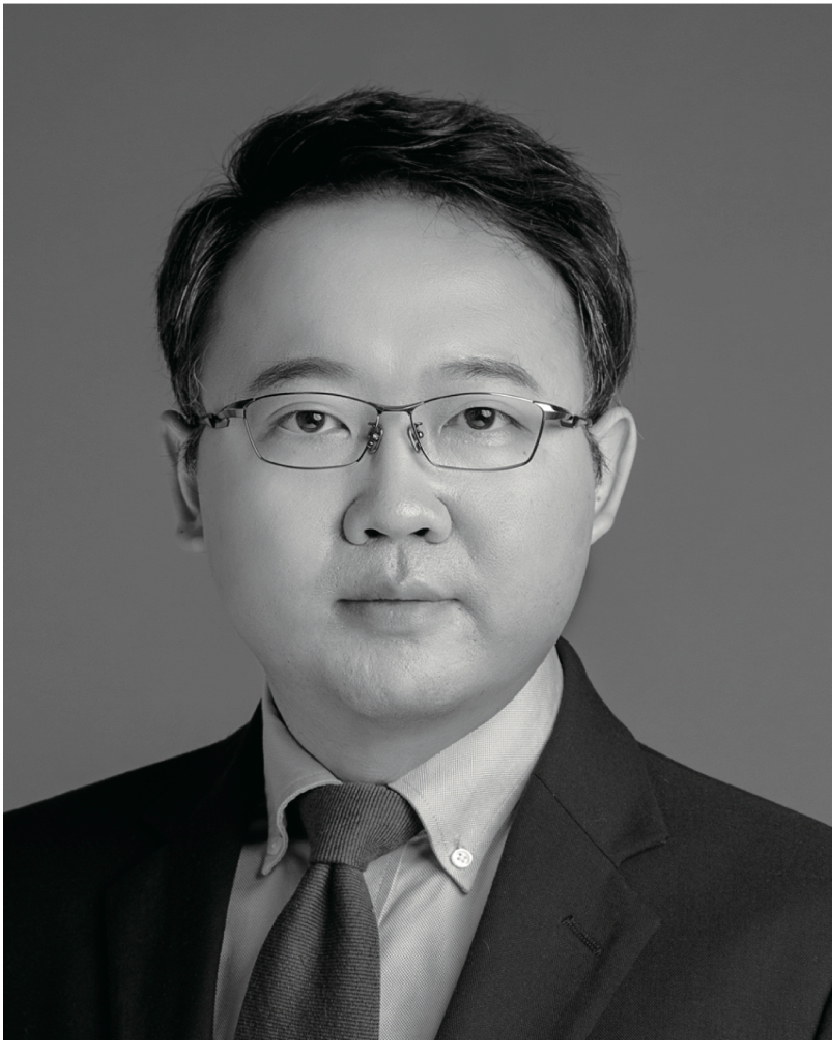}}]{Nan Cao} received his Ph.D. degree in Computer Science and Engineering from the Hong Kong University of Science and Technology (HKUST), Hong Kong, China in 2012. He is currently a professor at Tongji University and the Associate Dean of the Tongji College of Design and Innovation. He also directs the Tongji Intelligent Big Data Visualization Lab (iDV$^x$ Lab) and conducts interdisciplinary research across multiple fields, including data visualization, human computer interaction, machine learning, and data mining. Before his Ph.D. studies at HKUST, he was a staff researcher at IBM China Research Lab, Beijing, China. He was a research staff member at the IBM T.J. Watson Research Center, New York, NY, USA before joining the Tongji faculty in 2016.

\end{IEEEbiography}

\end{document}